\documentclass[11pt]{article}
\usepackage{jheppub}

\usepackage{epsfig}
\usepackage{amssymb}
\usepackage{amsfonts}
\usepackage{amsbsy}
\usepackage[all,v2,cmtip,2cell]{xy}
\usepackage{amsmath}
\usepackage{dsfont}
\usepackage{bbm}
\usepackage{upgreek}
\usepackage{amssymb,amscd}
\usepackage{graphicx}
\usepackage{mathrsfs}
\usepackage{amsmath,amsthm}
\usepackage{slashed}
\usepackage{dsfont}
\usepackage{tikz}
\usepackage[utf8]{inputenc}
\makeatletter
\DeclareFontFamily{OMX}{MnSymbolE}{}
\DeclareSymbolFont{MnLargeSymbols}{OMX}{MnSymbolE}{m}{n}
\SetSymbolFont{MnLargeSymbols}{bold}{OMX}{MnSymbolE}{b}{n}
\DeclareFontShape{OMX}{MnSymbolE}{m}{n}{
    <-6>  MnSymbolE5
   <6-7>  MnSymbolE6
   <7-8>  MnSymbolE7
   <8-9>  MnSymbolE8
   <9-10> MnSymbolE9
  <10-12> MnSymbolE10
  <12->   MnSymbolE12
}{}
\DeclareFontShape{OMX}{MnSymbolE}{b}{n}{
    <-6>  MnSymbolE-Bold5
   <6-7>  MnSymbolE-Bold6
   <7-8>  MnSymbolE-Bold7
   <8-9>  MnSymbolE-Bold8
   <9-10> MnSymbolE-Bold9
  <10-12> MnSymbolE-Bold10
  <12->   MnSymbolE-Bold12
}{}

\let\llangle\@undefined
\let\rrangle\@undefined
\DeclareMathDelimiter{\llangle}{\mathopen}%
                     {MnLargeSymbols}{'164}{MnLargeSymbols}{'164}
\DeclareMathDelimiter{\rrangle}{\mathclose}%
                     {MnLargeSymbols}{'171}{MnLargeSymbols}{'171}
\makeatother

\def\be{ \begin{equation} }
\def\ee{ \end{equation}}

\newcommand{\eq}[1]{\begin{align}\begin{split}#1\end{split}\end{align}}

\def\exp{{\rm exp}}

\def\half{\frac{1}{2}}

\def\one{{\hbox{ 1\kern-.8mm l}}}
\def\CA{{\cal A}}

\def\CH {{\cal H}}

\def\CN {{\cal N}}

\def\CP {{\cal P}}

\def\CH {{\cal H}}

\def\CT {{\cal T}}
\def\CU {{\cal U}}

\def\IC{\mathbb{C}}

\def\ICP{\mathbb{CP}}

\def\IR{{\mathbb{R}}}

\def\IZ{{\mathbb{Z}}}

\def\rmk#1{\bigskip\noindent{\bf Remark} }
\def\cnj#1{\bigskip\noindent{\bf Conjecture:} }
\def\rslt{\bigskip\noindent{\bf Result:} }

\DeclareMathAlphabet{\mathpzc}{OT1}{pzc}{m}{it}

\title{Anomaly Enforced Gaplessness and Symmetry Fractionalization for  $Spin_G$ Symmetries}

\author{T.~Daniel Brennan}

\affiliation{Department of Physics, University of California San Diego,\\
 \textit{9500 Gilman Drive, La Jolla CA 92093-0319, USA}}

\emailAdd{tbrennan@ucsd.edu}

\abstract{Symmetries and their anomalies give strong constraints on renormalization group (RG) flows of quantum field theories. Recently, the identification of a theory's global symmetries with its topological sector has provided additional constraints on RG flows to symmetry preserving gapped phases due to mathematical results in category and topological quantum field theory. In this paper, we derive constraints on RG flows from $\IZ_2$-valued pure- and mixed-gravitational anomalies that can only be activated on non-spin manifolds. We show that such anomalies cannot be matched by a unitary, symmetry preserving gapped phase without symmetry fractionalization. In particular, we discuss examples that commonly arise in $4d$ gauge theories with fermions. 
}

\begin{document}

\maketitle

\section{Introduction}

Symmetry is one of the guiding principles of modern physics. It can be used to derive selection rules for correlation functions, constrain non-perturbative dynamics, and give insight into renormalization group flows. Recently, the notion of symmetry groups has been expanded to the more general framework of symmetry categories. In the setting of physical theories, this categorical structure is realized via topological operators and their rules for braiding and fusion. This is the notion of ``generalized global symmetries'' or  ``categorical global symmetries.'' For a review see \cite{Cordova:2022ruw,Freed:2022qnc,Gaiotto:2014kfa,Schafer-Nameki:2023jdn,Brennan:2023mmt,Bhardwaj:2023kri,Shao:2023gho} and references therein.

This modern perspective on global symmetries has given us new tools to analyze quantum field theories. The fact that symmetries are realized by topological operators means that  symmetries of quantum field theories are classified by topological field theories which themselves are described by certain types of categories -- which (at least in principle) can be completely classified. 

In this paper we will restrict our attention to the case of higher form global symmetries which are characterized by their group-like structure. Higher form global symmetries can be decomposed into simple factors which are labeled by a degree $p\in\IZ_{\geq0}$ and a group $G^{(p)}$: each simple factor is referred to as a $p$-form global symmetry. A $p$-form global symmetry group $G^{(p)}$ can be either discrete or continuous  and acts on $p$-dimensional operators via co-dimension-$(p+1)$ topological operators called symmetry defect operators.

One of the results of axiomatizing symmetries in this way is that we are able to better characterize how a symmetries can be matched along their RG flows. 
The general picture is the following. When a theory has a symmetry group $G$, the Hilbert space transforms under (possibly projective) representations of $G$. Such a  theory may flow at low energies to a gapped or gapless theory -- i.e. the low energy spectrum of the Hamiltonian may be discrete or a continuous near zero. At long distances, the gapless phases are described by scale invariant theories while gapped phases are described by topological theories. 

 In addition, we can further distinguish theories by whether the symmetry $G$ is preserved or spontaneously broken. Such a symmetry $G$ is preserved if the vacuum state transforms in a trivial representation whereas it is broken if the vacuum state transforms in a non-trivial representation of $G$. The breaking of $G$ is indicated by the existence of charged operators that have a non-trivial vacuum expectation value. In the case where a symmetry is spontaneously broken, the theory is in general nontrivial: 
\begin{itemize}
\item A spontaneously broken 0-form symmetry $G^{(0)}$ leads to a degenerate ground state corresponding to the elements of the group -- for $G^{(0)}$ continuous, this degeneracy is encoded in a massless $G^{(0)}$-valued scalar Goldstone boson. 
\item A spontaneously broken $p>0$-form global symmetry does not necessarily have a degenerate Hilbert space on any manifold, but rather leads to a non-trivial Hilbert space on $T^p\times \IR^{d-p}$ corresponding to wrapped $p$-dimensional line operators which take non-trivial vevs.
\end{itemize}

Another possible behavior of symmetries along an RG flow is that it may develop a non-trvial categorical structure.\footnote{In fact, the emergence of non-trivial categorical structure along an RG flow can be used to infer the behavior of the theory at intermediate points along the flow \cite{Brennan:2020ehu,Cordova:2018cvg,Brennan:2023kpw,Cordova:2022ieu,Choi:2022jqy,Choi:2022rfe,Choi:2022fgx}.} 
One of the simplest ways a non-trivial categorical structure can emerge is by \emph{symmetry fractionalization} which for our purposes describes the case where a 0-form symmetry mixes with a (potentially emergent) 1-form symmetry \cite{Delmastro:2022pfo,Brennan:2022tyl,Barkeshli:2014cna,Chen:2014wse}.\footnote{More generally, symmetry fractionalization can describe when a $p$-form global symmetry can mix with a $(p+1)$-form global symmetry, but we will not need the general case here. For the more details see \cite{Delmastro:2022pfo,Bartsch:2023pzl,Bhardwaj:2023wzd,Bhardwaj:2022dyt}. }
This mixing can be realized by a choice 
 of $\alpha\in H^2(BG^{(0)},G^{(1)})$, which defines a central extension of $G^{(0)}$ by $G^{(1)}$, together with a 
modification of the product law of the $0$-form symmetry defect operators $U_g(\Sigma_{d-1})$:\footnote{
Note that a more general action of $G^{(0)}$ on the line operators that are protected by $G^{(1)}$-symmetry can be defined by picking an action $\rho:G^{(0)}\to Aut(G^{(1)})$ together with a twisted cohomology group $\alpha_\rho\in H^2_\rho(BG^{(0)},G^{(1)})$. This is described by a ``split  2-group'' (also sometimes called ``strict 2-group'') which will not be important for our discussion. See \cite{Bartsch:2023pzl,Bhardwaj:2023wzd,Baez:2008hz} for more details.  
}
\eq{
U_{g_1}(\Sigma_{d-1})\cdot U_{g_2}(\Sigma_{d-1})=\CU_{\alpha(g_1,g_2)}(\partial\Sigma_{d-1}) \,U_{g_1g_2}(\Sigma_{d-1})~, }
where $ g_1,g_2\in G^{(0)}$  and $\CU_h(\Gamma_{d-2})$ are 1-form symmetry defect operators for $h\in G^{(1)}$. Physically, the non-trivial $G^{(0)}$ action can be realized via a world volume anomaly for the line operators which can indicate that they describe the world line of some heavy particle that transforms projectively under $G^{(0)}$. \\

The IR behavior of QFTs are highly constrained by `t Hooft anomalies of the global symmetry group $G$. These anomalies must be matched along symmetry preserving RG flows and can used rule out different possibilities for the IR theory. 
These ideas go under the name of ``Lieb-Schultz-Mattis type theorems''  or ``anomaly enforced gaplessness'' 
 \cite{Lieb:1961fr,Wang:2014lca,Wang:2016gqj,Sodemann:2016mib,Wang:2017txt,Kobayashi:2018yuk,Cordova:2019bsd,Cordova:2019jqi,Apte:2022xtu,Wang:2017loc,Wan:2019oyr,Wan:2018djl}. 
For discrete symmetries in continuum QFTs, it was shown in \cite{Cordova:2019bsd,Cordova:2019jqi}, that  $d$-dimensional unitary theories with a global symmetry group $G$ that has an anomaly given by $\omega_{d+1}(A)\in H^{d+1}(BG;U(1))$, cannot be matched in the IR by a symmetry preserving gapped phase if the anomaly evaluates non-trivially on the mapping class torus 
\eq{
{\rm exp}\left\{2\pi i \int_{N_{d+1^{(\varphi)}}} \omega_{d+1}(A)\right\}\neq 1~,
}
where $N_{d+1}^{(\varphi)}=\big(S^p\times S^{d-p}\big)\rtimes_\varphi S^1$ and $\varphi$ is the twist by some $\varphi\in G$.

Recently, there has been interest in computing and understanding anomalies that are only activated on non-spin manifolds \cite{Wang:2018qoy,Brennan:2023mmt,BI,Ang:2019txy,Wang:2017loc,Wan:2019oyr,Wan:2018djl}. These anomalies are more subtle to compute than their perturbative counter parts, often requiring more sophisticated index theorem computations.  

In this paper, we will consider the anomalies of theories which have symmetries that are of the form 
\eq{
G_{\rm total}=\frac{G\times Spin(d)}{\IZ_2}~,
}
where $G$ is the internal global symmetry group, $Spin(d)$ is the $d$-dimensional Euclidean Lorentz group, and the $\IZ_2$ quotient identifies $(-1)^F$ with a (potentially trivial) $\IZ_2$ subgroup of the center of the 0-form subgroup of $G$. In other words, the global symmetry group $G_{\rm total}$ corresponds to taking the quotient of $G\times Spin(d)$ by a $\IZ_2$ 0-form symmetry generated by $(-1)^F$ or $(-1)^F\circ (-\mathds{1}_G)$ which acts trivially on the gauge invariant local operators. 
 When the $\IZ_2$ quotient is generated by $(-1)^F$, we call the theory bosonic and say that the theory has bosonic symmetry, whereas when the $\IZ_2$ quotient is generated by $(-1)^F\circ (-\mathds{1}_G)$, we say the symmetry has a $Spin_G$-symmetry. Here we will treat bosonic symmetries as a special case of $Spin_G$-symmetries with trivial $G$-structure -- i.e.  $G$ is the trivial group. 

Theories with  $Spin_G$-symmetry can be put on non-spin manifolds:\footnote{
More precisely, the theory can be put on manifolds with $Spin_G$ structure. For the case where $G$ is abelian, a $Spin_G$ connection is usually referred to as a $Spin_\IC$ connection and the obstruction to a $Spin_\IC$ structure is given by the third Stiefel-Whitney class of the spacetime manifold $w_3(TM)$. See \cite{Wang:2018qoy} for further discussion about obstructions to $Spin_G$-structures for $G$ non-abelian. 
} this requires turning on a $\IZ_2$-discrete flux in $G$ for the case with non-trivial $G$-structure. This allows for new classes of $\IZ_2$-valued pure- and mixed-gravitational anomalies whose anomalous phase couples to the Stiefel-Whitney class $w_2(TM)$. Another important feature of $Spin_G$-symmetries is that the $\IZ_2$ quotient in $G_{\rm total}$ implies $ H^2(BG_{\rm total},\IZ_2)\neq 0$ which allows for symmetry fractionalization. The symmetry fractionalization is realized by line operators that transform non-trivially under a choice of integral lift of $w_2(TM)+w_2(G)$ for $Spin_G$-symmetries which describes the world-line of a heavy particle that transforms non-trivially under $(-1)^F\circ(-\mathds{1}_{G})$.  In the special case of bosonic symmetries, the line operators transform non-trivially under a choice of integral lift of $w_2(TM)$ which describes the world-line of a heavy fermion. 

In this paper, we will derive new constraints on the IR phase of a theory with anomalies of such $Spin_G$-symmetries. Many anomalies of theories with and $Spin_G$-symmetries can be analyzed on mapping class tori of a product of spheres: such anomalies lead to constraints on the IR phase due to the theorem of \cite{Cordova:2019bsd,Cordova:2019jqi} described above. However, there also exist anomalies which vanish on spin manifolds that do not lead to such constraints. Here we will prove that such anomalies that are only activated on non-spin manifolds do lead to constraints on RG flows. In particular, such anomalies imply that the theory is either gapless, experiences symmetry fractionalization, or that the symmetry group $G$ is spontaneously broken in the deep IR.

%\cg{Can do higher $G^{(p)}$ on $\ICP^{p-1}$}
Our anomaly can be stated more formally as follows. 
Here we will consider a $d$-dimensional theory $\CT_d$ for $d= 4,6$  on a space-time manifold $M$ with internal symmetry group $G=G_0\times G_1=G_0\times \left(G^{(0)}\times G^{(1)}\right)$  and  $Spin_G$-symmetry structure. Here $G_0$ is the part of $G$ that participates in the $Spin_G$ structure -- i.e. the $Spin_G$ structure can be reduced to a $Spin_{G_0}$ structure -- while $G_1=G^{(0)}\times G^{(1)} $ is a product of discrete higher form symmetry groups. Here we make no assumption of whether or not $G_0$ is discrete or continuous. 

We turn on background gauge fields for the symmetry group $G_1$ by coupling the theory to a non-trivial bundle $P\otimes TM$ with connection where $P$ is a principal $G_1$-bundle and $TM$ is the tangent bundle of $M$. When $M$ is a non-spin manifold, we must elevate the tangent bundle to a $Spin_{G_0}$ bundle which requires an additional $G_0/\IZ_2$ bundle with fixed non-trivial obstruction class $w_2(G_0)=w_2(TM)$. 
We will denote the background gauge fields for $G_1$ by $\{A\}$ and will assume that $G$ has anomalies described by the $(d+1)$-dimensional anomaly polynomial $\omega_{d+1}(A;w_2(TM))$ is a $U(1)$-valued characteristic class of $P\otimes TN$ where $ N$ is a $(d+1)$-manifold.  We assume that $\omega_{d+1}(A;0)=0$ -- i.e. that the anomaly vanishes on spin manifolds for any choice of $G_1$-bundle. 

\rslt \emph{Consider a $d$-dimensional QFT with $Spin_G$-symmetry and anomaly $\omega_{d+1}(A;w_2(TM))$ as above. Define the $(d+1)$-dimensional manifold $N_{d+1}^{(g)}$ which is topologically the product manifold  $N_{d+1}^{(g)}=\ICP^2\times S^{p}\times S^1$ where $p=d-4$ we twist the fibers of $P\otimes TN$ by an  element $g\in G$ along the $S^1$-direction which we write as a mapping class torus:}
\eq{
N_{d+1}^{(g)}=(\ICP^2\times S^{p})\rtimes_g S^1~.
}
\emph{If the anomaly of a theory on $N_{d+1}^{(g)}$ can evaluate to a non-trivial phase:}
\eq{
{\rm exp}\left\{2\pi i \int_{N_{d+1}^{(g)}} \omega_{d+1}(A;w_2(TN))\right\}\neq 1~,
}
\emph{then the theory $\CT_d$ cannot flow to a symmetry preserving gapped phase without symmetry fractionalization for $(-1)^F$ (which for $Spin_G$-symmetries is identified with an element of $G$). }\\

 This result generalizes those of \cite{Cordova:2019bsd,Cordova:2019jqi} to include $Spin_G$ global symmetries at the expense of a weaker constraint on the IR behavior. \\

The outline of our paper is as follows. We begin by reviewing the results of \cite{Cordova:2019bsd,Cordova:2019jqi} which forms the cornerstone of our analysis. In particular, we review how anomalies evaluated on the mapping class torus of a product of spheres can imply that the theory cannot flow to a symmetry preserving gapped phase. In Section \ref{sec:nonspin} we prove our main result by adapting the analysis of \cite{Reutter:2020bav} to the case of non-spin mapping class tori.

We conclude with a set of examples. Of particular interest are the anomalies of $4d$ theories. Here, we will discuss two types of anomalies that are sensitive to the $(-1)^F$ symmetry: 
\eq{\label{4dexamplesintro}
\CA_1=\pi i \int x_1\cup y_2\cup w_2(TM)\quad, \quad \CA_2=\pi i \int w_2(TM)\cup z_3
}
where $x_1,y_2,z_3$ are degree -1,-2, and -3   $\IZ_2$-valued cohomology classes respectively. As we will discuss, these anomalies can be matched by flowing to a phase where the $(-1)^F$ symmetry is fractionalize. Such anomalies often arise in $4d$ gauge theories  such as  2-flavor $SU(N_c)$ adjoint QCD \cite{BI,Donaldson:1983wm,Donaldson:1990kn,Witten:1988ze,Witten:1994cg,Moore:1997pc}. The results of our paper can be used to infer characteristics of the IR limit of these theories.

\section{Review of Anomaly Enforced Gaplessness on Spin Manifolds}

\label{sec:spin}

As we have discussed, the derivation of ``anomaly enforced gaplessness'' for theories that have anomalies for discrete symmetries on spin manifolds 
is given in \cite{Cordova:2019bsd,Cordova:2019jqi}.  
The main result  
can be stated as follows: 

\bigskip\noindent{\bf Result of \cite{Cordova:2019bsd,Cordova:2019jqi}:} Consider a $d$-dimensional  theory $\CT_d$ with a discrete, abelian symmetry group $G$ which has an anomaly polynomial given by $\omega_{d+1}(A)\in H^{d+1}(BG,U(1))$. Let us consider turning on background gauge fields for $G$, which we denote $\{A\}$ -- i.e. we couple to a principal $G$ bundle $P$. Take the $(d+1)$-dimensional manifold $N_{d+1}=S^{p+1}\times S^{d-p-1}\times S^1$ where we additionally twist $P$ along the $S^1$ factor by an element $g\in G$: $N_{d+1}^{(g)}=(S^{p+1}\times S^{d-p-1})\rtimes_g S^1$. If 
\eq{
{\rm exp}\left\{2\pi i \int_{N_{d+1}^{(g)}}\omega_{d+1}(A)\right\}\neq 1~, 
}
then there is no unitary, symmetry preserving TQFT with anomaly $\omega_{d+1}(A)$. In particular, this means that the theory cannot flow to a symmetry preserving gapped unitary TQFT. \\
~\\

\noindent This result can be proven as follows.  \\

Consider a $d$-dimensional unitary TQFT. First note that unitary TQFTs are by definition reflection positive \cite{Freed:2016rqq}. This means that for any $d$-manifold $X_d=\overline{Y_d}\#Y_d$ that $Z[X_d]\geq0$ where $\overline{Y_d}$ is the orientation reversal of $Y_d$. This can be seen from the fact that the path integral on $Y_d$ defines a state $\big{|}Y_d\big\rangle$ which is an element of the Hilbert space $\CH_{\partial Y_d}$ defined on $\partial Y_d$ which allows us to write 
\eq{
Z[X_d]=\big\langle Y_d\big{|}Y_d\big\rangle\geq0~,
}
by unitarity. In particular, any unitary TQFT has the condition that $Z_{\rm TQFT}\big[S^d\big]> 0$. 
Using the fact that $S^d$ can be decomposed by cutting along the equator, $S^d=\overline{D^d}\#D^d$ 
where $D^q$ is the $q$-dimensional disk, 
we see that  
%that the theory has a well defined  have $Z_{\rm TQFT}\big[S^d\big]>0$ which 
%Since the vacuum state is prepared by evaluating the path integral on $D^d$, which in a unitary TQFT is a non-trivial state, we have the condition
\eq{
Z\big[S^d\big]=\big\langle 0\big{|}0\big\rangle~. 
}
We  will choose to normalize the vacuum state $|0\rangle$ (or equivalently add counter terms to the action) such that $Z_{\rm TQFT}\big[S^d\big]=1$. 

Additionally, in a unitary TQFT, the Hilbert space on $S^{d-1}$ is completely generated by the vacuum state \cite{Reutter:2020bav}: 
\eq{
\CH\big[S^{d-1}\big]\cong \IC\big[\big{|}0\big\rangle\big]~.
}
This follows from the fact that there is a unique local operator that is topologically invariant -- the identity operator -- which on $D^d$ generates the vacuum state.  

We can additionally cut $S^d$ along $S^{d-p-1}\times S^p$:
\eq{
S^d=\overline{M_1}\# M_2:=\Big(\overline{D^{d-p}\times S^p}\Big)\#\Big(S^{d-p-1}\times D^{p+1}\Big)~.
}
 This means that for any TQFT the states created by evaluating the path integral on $M_1,M_2$ are both non-trivial:
\eq{
Z_{\rm TQFT}\big[S^d\big]=\big\langle 0\big{|}0\big\rangle=Z_{\rm TQFT}[\overline{M_1}\#M_2]=\big\langle M_1\big{|}M_2\big\rangle=1~. 
}
This further means that 
\eq{
Z_{\rm TQFT}[S^{d-p-1}\times S^{p+1}]=\big\langle M_2\big{|}M_2\big\rangle>0
}
and analogously for $M_1$.

We will also need the fact that if the internal global symmetry group $G$ is not spontaneously broken, the partition function on $S^{d-p-1}\times S^{p+1}$ is non-zero  for non-trivial background gauge fields:
\eq{
Z[S^{d-p-q}\times S^{p+1};A]\neq 0~. \label{positivitybackgroundPF}
}
For the moment, we will assume this -- we will reproduce the proof from \cite{Cordova:2019bsd} in the next section for completeness. 

Now let us analyze a QFT that has an anomaly is that described by a $(d+1)$-dimensional SPT phase  $\omega_{d+1}(A)\in H^{d+1}(BG,U(1))$. The phase that the partition function picks up under a constant background gauge transformation defined by $g\in G$:  $A\mapsto A^{(g)}=A$ is identical to the phase produced by the anomaly polynomial on the product manifold $N_{d+1}^{(g)}=\big(S^{d-p-1}\times S^{p+1}\big)\rtimes_g S^1$ where we impose twisted boundary conditions for the principal $G$-bundle on $S^{d-p-1}\times S^{p+1}$ along the $S^1$: 
\eq{
Z_{\rm QFT}\Big[S^{d-p-1}\times&S^{p+1};A^{(g)}\Big]=\exp\left\{i \int_{N_{d+1}^{(g)}}\omega_{d+1}(A)\right\}\times Z_{\rm QFT}\Big[S^{d-p-1}\times S^{p+1};A\Big]~,
}
where $\omega_{d+1}(A)$ is integrated over $N_{d+1}^{(g)}$.  
Since the twist $g$ is a flat background transformation which acts trivially on the gauge field for discrete abelian symmetries,  
 we see that either 
\eq{
{\rm exp}\left\{i \int_{N_{d+1}^{(g)}}\omega_{d+1}(A)\right\}=1\quad{\rm or} \quad Z_{\rm QFT}[S^{d-p-1}\times S^{p+1},A]=0~.
} 

\noindent For a TQFT, the fact that $g$ acts trivially together with \eqref{positivitybackgroundPF} implies that the anomaly must satisfy
\eq{\label{TQFTcond}
{\rm exp}\left\{i \int_{N_{d+1}^{(g)}}\omega_{d+1}(A)\right\}=1~.
} 
This means that if 
\eq{
{\rm exp}\left\{i \int_{N_{d+1}^{(g)}}\omega_{d+1}(A)\right\}\neq 1\quad\Longrightarrow \quad Z_{\rm QFT}\Big[S^{d-p-1}\times S^{p+1};A\Big]=0~.
}
Since we know that for a unitary, symmetry preserving TQFT, $Z_{\rm TQFT}\Big[S^{d-p-1}\times S^{p+1};A\Big]\neq0$, we must have that the IR theory cannot be matched by such a TQFT, and hence the theory cannot flow to a gapped phase.   

\subsection{Proof of \eqref{positivitybackgroundPF}}

\label{sec:lemma}

Now we will prove $Z[S^{d-p-1}\times S^{p+1}]\neq 0$ implies that $Z[S^{d-p-1}\times S^{p+1};A]\neq 0$ if the associated internal global symmetry is not spontaneously broken. We will do this in steps. First, we will prove the following Lemma. \\

\noindent \textbf{Lemma:}~\emph{Let us consider a symmetry defect operator for a symmetry group $G$ wrapped on $S^{d-p}$: $U_g\Big(S^{d-p-1}\times \{\rm pt\}\Big)$ and let us assume that $G$ is not spontaneously broken. 
Then, the symmetry defect operator can be replaced by wrapping a corresponding symmetry defect operator on a disk $D^{d-p-1}$ where we additionally impose a boundary condition $\phi$ on the boundary: }
\eq{
\big\langle U_g\Big(S^{d-p-1}\times \{{\rm pt}\}\Big)...\big\rangle= \big\langle U_g^{(\phi)}\Big(D^{d-p-q}\times\{{\rm pt}\}\Big)...\big\rangle~.
}

 \bigskip\noindent\textit{pf.}   First, if a $p$-form symmetry defect operator $U_g$ corresponds to a preserved symmetry $G^{(p)}$, then $U_g$ has trivial winding with any non-trivial operator on $S^{d-p-1}\times S^{p+1}$.  
This follows from the assumption that $G$ is preserved, as this implies no charged operator has a non-trivial expectation value.  
Now to show that a symmetry defect operator can end on a topological operator $\phi$, we can use the state operator correspondence to equate existence of $\phi$ to the fact that the boundary state $|\phi_g\rangle$ which is created 
from cutting $U_g(S^{d-p-1})$ along a great circle:
\eq{
\langle U_g(S^{d-p-1})\rangle_{S^{p+1}\times S^{d-p-1}}=\langle \phi_g|\phi_g\rangle ~,
}
is non-null state. This state belongs to the twisted Hilbert space $|\phi_g\rangle\in \CH[S^{p+1}\times S^{d-p-2}_g]$. As it turns out, we can construct a linear map 
\eq{
f:\CH[S^{p+1}\times S_g^{d-p-2}]\to \CH[S^{d-1}_g]~, 
}
where $f$ is the cobordism between $S^{p+1}\times S^{d-p-2}\to S^{d-1}$. This can be explicitly constructed as in \cite{Cordova:2019bsd} by excising $D^{d-p-1}\times S^{p+1}$ from $D^p$ where we wrap a $U_g$ on a $S^{d-p-2}\times [0,1]$ between the two boundary components. Since $f$ is linear and $f:|\phi\rangle\mapsto |0_g\rangle$ where $|0_g\rangle$ is the generator of $\CH[S^{d-1}_g]\cong \IC$ which by definition is a non-trivial state, we see that $|\phi_g\rangle$ is also a non-trivial state and therefore that there exists a $(d-p-2)$-dimensional operator $\phi$ on which $U_g$ can end. 
%
%We can show this by constructing a linear map from 
%
%
%where 
%\eq{
%\big{|}U_g\big\rangle:=Z[D^{d-p-1}_{U_g}\times S^{p+1}]~,
%}
%is the path integral on $D^{d-p-1}\times S^{p+1}$ with a defect operator inserted on $D^{d-p-1}$ 
%and $\big{|}0_g\big\rangle$ is the state on $S^{d-1}$ created by evaluating the partition function on $D^d$ where $U_g$ is wrapped on a $D^{d-p-1}$-disk. 
%
%We can show that the state $\big{|}0_g\big\rangle$ is non-trivial by computing its norm: 
%\eq{
%\big\langle 0_g\big{|}0_g\big\rangle=\big\langle U_g(S^{d-p-1})\big\rangle_{S^d}=Z\big[S^d\big]=1~,
%}
%since $S^{d-p-1}\subset S^d$ is homologous to a point. 
%Further, since $\CH\big[S^{d-1}\big]=\IC\big[\big{|}0\big\rangle\big]$, $\big{|}0_g\big\rangle=\big{|}0\big\rangle$. This implies that 
%\eq{
%\big\langle U_g(S^{d-p-1}\times\{{\rm pt}\})\big\rangle=\big\langle 0_g\big{|}U_g\big\rangle=\big\langle 0\big{|}U_g\big\rangle=\big\langle U_g^{(\phi)}(D^{d-p-q}\times \{pt\})\big\rangle
%}
%and that we are  simply cut a hole in the symmetry defect operator. The argument follows similarly for correlation functions involving other operators as long as we cut out a $d$-ball surrounding a point on $U_g$ that is away from any charged operators. While we have not explicitly constructed the boundary condition $\phi$, this proves the Lemma. 
For more details see \cite{Cordova:2019bsd,Cordova:2019jqi}. \qed

\bigskip Using this Lemma, we can prove \eqref{positivitybackgroundPF}. Consider the partition function $Z[S^{d-p-1}\times S^{p+1};A]$ where the background gauge fields are non-trivial along $S^{d-p-1}$ and $ S^{p+1}$. This is equivalent to wrapping symmetry defect operators   $U_g$ on $S^{p+1}$ and $\tilde{U}_{g^\prime}$ on $S^{d-p-1}$:
\eq{
Z[S^{d-p-1}\times S^{p+1};A]=\big\langle U_g(S^{d-p-2}\times\{{\rm pt}\})\, \widetilde{U}_{g^\prime}(\{{\rm pt}\}\times S^{p+1})\big\rangle~.
}
 Using the lemma, we can cut these symmetry defect operators 
 \eq{
 Z[S^{d-p-1}\times S^{p+1};A]=\big\langle U_g^{(\phi)}(D^{d-p-2}\times\{{\rm pt}\})\, \widetilde{U}_{g^\prime}^{(\tilde\phi)}(\{{\rm pt}\}\times D^{p+1})\big\rangle~,
 }
 and by their topological nature, contract them to an infinitesimally small ball. 
  We can then cut out a $D^d$ surrounding the contracted symmetry defect operators and replacing it with a boundary condition $\Phi$ which corresponds to a state $\big{|}\Phi\big\rangle$ in the Hilbert space on $S^{d-1}$: $\CH\big[S^{d-1}\big]$:
  \eq{
   Z[S^{d-p-1}\times S^{p+1};A]=\big\langle 0\big{|}\Phi\big\rangle~.
  }
However, following from our previous discussion, unitarity implies   that the state $\big{|}\Phi\big\rangle$ is proportional to the identity state $\big{|}\Phi\big\rangle\sim \big{|}0\big\rangle$. We can additionally see that this is non-zero, by computing its norm 
\eq{
\big\langle \Phi\big{|}\Phi\big\rangle=\big\langle U_g(S^{d-p-2}\times\{{\rm pt}\})\, \widetilde{U}_{g^\prime}(\{{\rm pt}\}\times S^{p+1})\big\rangle_{S^d}=\big\langle \mathds{1}\big\rangle_{S^d}=Z\big[S^d\big]=1~,
}
where here we used the fact that $S^{d-p-2}$ and $S^{p+1}$ are contractible in $S^d$. 
Now, since $\big{|}\Phi\big\rangle\propto\big{|}0\big\rangle$, we see that 
\eq{
Z[S^{d-p-1}\times S^{p+1};A]\propto Z[S^{d-p-1}\times S^{p+1}]\neq 0~,
}
thus proving  \eqref{positivitybackgroundPF}.

 \section{Anomaly Enforced Gaplessness for $Spin_G$ Symmetries}

\label{sec:nonspin}

A salient feature of the result of \cite{Cordova:2019bsd,Cordova:2019jqi}, which is reproduced in the previous section, is that all spheres and their products are spin manifolds. This means that such constraints are insensitive to $\IZ_2$-valued anomalies of $Spin_G$ symmetries that couple to the spin structure. Indeed, many $\IZ_2$-valued anomalies trivially evaluate on the product of spheres due to the fact that they are spin manifolds.  

To be clear, by $Spin_G$-symmetries we mean the following.   Consider a $d \geq 4$-dimensional theory that transforms faithfully under an internal symmetry group $G$. Here, we take $G$ to be a product of higher form symmetry groups 
\eq{\label{symgen}
G=G_0\times G_1=G_0\times \left(G^{(0)}\times G^{(1)}\right)~,
}
where we assume that $G_1$ is discrete and that $G_0$ is an additional 0-form global symmetry.  

We refer to $Spin_G$-symmetries as ones with a faithful symmetry group\footnote{It is also interesting to consider the case of higher group and non-invertible global symmetries in which $(-1)^F$ is identified with a component of the general symmetry structure. Here we will only focus on the case of higher group global symmetries and leave the more general story for future investigation.} 
\eq{ 
G_{\rm total}=\frac{G\times Spin(d)}{\IZ_2}~,
}
where the $\IZ_2$ acts as $(-1)^F$ in $Spin(d)$ and acts on $G_0\subset G$. We call  theories where the $\IZ_2$ acts trivially on $G$ as ``bosonic'' and their symmetries as bosonic symmetries. Bosonic symmetries are a special case of $Spin_{G_0}$-symmetry with trivial $G_0$ structure: where $G_0$ is the trivial group.  

Here the structure of the symmetry group allows us to put the theory on a non-spin manifold (with $Spin_G$ structure). Bosonic theories can be naturally put on non-spin manifolds. 
For the more general case where $\IZ_2$ acts on both $Spin(d)$ and $G$, the theory can only be put on a non-spin manifold by elevating the $G$-bundle to a $Spin_{G}$-bundle as described in \cite{Wang:2018qoy}. This restricts $G$ to $G/\IZ_2$ bundles with fixed obstruction class $w_2(G)\in H^2(BG/\IZ_2;\IZ_2)$ such that $w_2(TM)=w_2(G)$.  Here we will restrict our attention to the case where the $\IZ_2\subset G_0$ that participates in the $Spin_G$ structure is non-abelian and to manifolds that admit such $Spin_G$ structures. 

In the following analysis we will show how the previous section can be generalized to encompass the anomalies of $Spin_G$-symmetries that are only activated on non-spin manifolds.   
In order to generalize the previous results, it is important to highlight what elements of the previous derivation  relied on evaluating the anomaly on the mapping class torus of a product of spheres. As it turns out, the only (independent) statement that  relied on the exact topology of the product of spheres was to show that $Z[S^{p+1}\times S^{d-p-1}]\neq 0$. 

Thus, in order to probe the anomalies that are only activated on non-spin manifolds, we need to find a $d$-dimensional non-spin manifold $M_d$ such that 
\eq{
Z[M_d]\neq 0~.
}
Given such a manifold, the other steps of proving anomaly enforced gaplessness proceed as in the case of $S^{d-p-1}\times S^{p+1}$. Using the result from Section \ref{sec:lemma}, one can then show that if $G$ is not spontaneously broken, then $Z_{TQFT}[M_d;0]\neq 0$ implies $Z_{TQFT}[M_d;A]\neq 0$, and then use the mapping class argument to show that 
\eq{
Z_{QFT}[M_d;A]={\rm exp}\left\{i \int_{N_{d+1}^{(g)}}\omega_{d+1}(A)\right\}Z_{QFT}[M_d;A]
}
which again would imply that either 
\eq{
{\rm exp}\left\{i \int_{N_{d+1}^{(g)}}\omega_{d+1}(A)\right\}=1\quad {\rm or}\quad Z_{QFT}[M_d]=0~.
}
Therefore, the non-trivial anomaly would imply that the theory is gapless or that $G$ is spontaneously broken up to other possible assumptions.

\bigskip It then remains to pick an appropriate non-spin manifold $M_d$ such that  under some conditions $Z_{TQFT}[M_d]\neq 0$ for unitary TQFTs.  

\subsection{$Z_{TQFT}$ on non-Spin Manifolds}
 
 A particularly nice choice of $M_d$ non-spin is 
 \eq{
 M_d=\ICP^2\times S^{p}~. 
 }
 As we will show in this subsection, for any unitary TQFT, $Z_{TQFT}[\ICP^2\times S^{p}]\neq 0$ when we turn on appropriate fluxes for the $Spin_G$-structure.  \\

\noindent First we will need the following proposition  from \cite{Reutter:2020bav}.

\bigskip\noindent\textbf{Proposition:} \emph{For a unitary $d$-dimensional TQFT, then the partition function  $Z$ is multiplicative under connected sums: } 
\eq{\label{prop}
Z[M\#N]= Z[M]\,Z[N]~.
}

\bigskip\noindent\textit{pf.} Under the decomposition 
\eq{
M\#N=\big(M\backslash D^d\big)\bigcup_{S^{d-1}}\big(N\backslash D^d\big)
}
we can write the partition function of the connected sum as 
\eq{
Z\big[M\#N\big]=Z\big[M\backslash D^d\big]Z\big[N\backslash D^d\big]=\big\langle \overline{M\backslash D^d}\big{|}N\backslash D^d\big\rangle~.
}
Now since $\CH\big[S^{d-1}\big]\cong \IC\left[\big{|}D^d\big\rangle\right]$, we can insert a complete set of states:
\eq{
Z\big[M\#N\big]&=\Big\langle \overline{M\backslash D^d}\Big{|}\frac{\big{|}D^d\big\rangle \big\langle D^d\big{|}}{\big\langle D^d\big{|}D^d\big\rangle}\Big{|}N\backslash D^d\Big\rangle=\frac{1}{Z\big[S^d\big]}\big\langle \overline{M\backslash D^d}\big{|}D^d\big\rangle\big\langle D^d\big{|}N\backslash D^d\big\rangle\\
&=\frac{Z\big[M\big]Z\big[N\big]}{Z\big[S^d\big]}=Z\big[M\big]\,Z\big[N\big]~. 
}
Thus proving the proposition. \qed

\subsection{Generalized Gluck Twist and Symmetry Fractionalization}

Now we would like to consider the case of a $4d$ TQFT on the manifold $\ICP^2\#\overline{\ICP^2}$ with the symmetry as in \eqref{symgen}.\footnote{Here, since we are considering a TQFT, all operators are protected by a corresponding symmetry. Our assumption that the theory has the symmetries given by \eqref{symgen} implies that there can only be symmetry defect operators of co-dimension $1,2$ which obey a group-like product rule, as well as possibly operators of dimension $0,1$ which are charged under these symmetries. } This manifold   can be obtained by taking  two copies of $ S^2\times D^2$ and gluing them together with a twist called the ``Gluck twist'' \cite{Glu62,Reutter:2020bav}:
\eq{
(\vec{x},t)\in S^2\rtimes S^1\sim (R_{z}(2\pi)\cdot \vec{x},t+2\pi)~.
}
This means that we can identify 
\eq{
Z\big[\ICP^2\# \overline\ICP^2\big]=\big\langle S^2\times D^2\big{|}\widehat{\CT}\big{|}S^2\times D^2\big\rangle~,
}
where $\widehat{\CT}$ is the ``Gluck twist operator'' that enacts the Gluck twist. We then  have the result that \underline{if} $\widehat{\CT}$ acts trivially, then 
\eq{
Z\big[\ICP^2\# \overline\ICP^2\big]=\big\langle S^2\times D^2\big{|}\widehat{\CT}\big{|}S^2\times D^2\big\rangle=\big\langle S^2\times D^2 \big{|}S^2\times D^2\big\rangle=Z\big[S^2\times S^2\big]>0~. 
}
We then see that using the proposition, that the property $Z\big[\ICP^2\# \overline\ICP^2\big]>0$ implies 
\eq{
Z\big[\ICP^2\#\overline\ICP^2\big]=\frac{1}{Z\big[S^4\big]}Z\big[\ICP^2\big]Z\big[\overline\ICP^2\big]>0\quad\Longrightarrow \quad Z\big[\ICP^2\big]\neq 0~. 
}

Let us be more precise about the meaning of the above statement. Consider a $4$-dimensional unitary TQFT coupled to a $Spin_G$-bundle: i.e. a $4d$ unitary TQFT with $Spin_G$-structure. Let us consider the Hilbert space on $S^2\times S^1$ which can be generated by evaluating the path integral on $S^2\times D^2$: $\CH[\partial(S^2\times D^2)]=\CH[S^2\times S^1]$. This Hilbert space is generated by the set of local operators and line operators wrapping $S^1$. Here the symmetry defect operators for $G_1$ (which are the only other non-trivial topological operators) do not generate non-trivial states in $\CH[S^2\times S^1]$ when $G_1$ is not spontaneously broken due to the fact that they have trivial linking with all operators as discussed in the previous section as discussed in Section \ref{sec:lemma} (see \cite{Cordova:2019bsd,Cordova:2019jqi} for more details). 

In terms of symmetries, the Gluck twist can be restated as a twist by $(-1)^F$ in the $S^2$ direction along the $S^1$. Because of this, $\widehat{\CT}$ will only act non-trivially on states that transform as a spinor on the $S^2$ factor. The operators that are acted on non-trivially by the twist are the local operators and the line operators wrapping the $S^1$. %The reason that the dimension 2 surface operators cannot carry a $(-1)^F$ world volume anomaly \cite{Bhardwaj:2022dyt,Delmastro:2022pfo}: i.e. there always exists a scheme in which the dimension 2 surface operators are bosonic. Similarly, for dimension 3 operators on $\partial(S^2\times D^2)$, there also exists a scheme in which the operator transforms trivial under the Gluck twist.\footnote{\color{red}{Are these operators consistent on non-spin manifold?} 
%To be more precise, the dimension 3 operators operators can have a 4-form world volume anomaly which can depend on the spin structure as $\CA_4=\pi i \int w_2(TM)\cup x_2$ where $x_2\in H^2(M;\IZ_2)$ is a $\IZ_2$-valued background gauge field. In general, one can pick a scheme where the surface operator only transforms under transformations of $x_2$. This makes the surface operator invariant under $(-1)^F$ transformations unless $x_2=w_2(TM)$. However, when $x_2=w_2(TM)$, the anomaly implies that the dimension 3 surface operator $\CD$ acquires a phase \eq{\label{CDphase}
%\widehat\CT \,\CD(S^2\times S^1)=e^{\pi i \oint_{S^2}w_2(TM)}\,\CD(S^2\times S^1)\,\widehat\CT~.
%}
%However, the $S^2$ embeds into $\ICP^2\#\overline{\ICP}^2$ as a cycle with $\oint_{S^2}w_2(TM)=0$ mod$_2$. One way to see this is to note that $\ICP^2=D^4\#(S^2\times D^2)$ can be realized as the fibering of a $S^1$ over an $S^2$ fibered over the interval. Here, the $D^2$ is glued into the Hopf fiber of the $S^3$ on the boundary of $D^4$. Because of this, the $S^1$ collapses at the origin of the $D^4$ and the origin of the $D^2$, forming the generator of $H_2(\ICP^2;\IZ)$ which is Pontryagin dual to $w_2(TM)$. Therefore the phase in \eqref{CDphase} is trivial and any dimension 3 surface operator on $\partial(D^2\times S^2)$ generates a bosonic state. 
%}

For bosonic theories, there are no local operators that transform non-trivially under $\widehat{\CT}$. However, for the theory with $Spin_{G}$-symmetry, the standard Gluck twist will  act non-trivially. The reason is that the structure of $\frac{G\times Spin(4)}{\IZ_2}$ in general identifies $(-1)^F$ with some generator $-\mathds{1}_G$ of a $\IZ_2\subset G$. By assumption, this quotient can be reduced to an action of $\IZ_2\subset G_0$ generated by $-\mathds{1}_{G_0}$. 
This means that there is some local operator that transforms non-trivially under $(-1)^F$, although it transforms trivially under $(-1)^F\circ (-\mathds{1}_{G_0})$. 
Instead,  for $Spin_{G}$-symmetries, 
we can construct the generalized Gluck twist operator $\widehat{\CT}_G$  which twists by the combined action of $(-1)^F\circ (-\mathds{1}_{G_0})$. 
Because the quotient acts at most on the 0-form component  $G_0\subset G$, there are no local operators that transform non-trivially under the combination of $(-1)^F\circ (-\mathds{1}_{G_0})$. 

Even though local operators do not transform under $\widehat{\CT}_G$ for $Spin_G$-symmetries or $\widehat{\CT}$ for  bosonic symmetries,  the line operators can transform non-trivially under  $(-1)^F\circ (-\mathds{1}_{G_0})$ or $(-1)^F$ respectively.  This is the statement that the $Spin_{G_0}$-symmetry can fractionalize \cite{Delmastro:2022pfo,Brennan:2022tyl,Barkeshli:2014cna,Chen:2014wse}. As we have discussed in the introduction, the relevant symmetry fractionalization for this setting is that line operators (which are usually only acted on by 1-form global symmetries) are acted on by a 0-form global symmetry which does not act on any local operators.  Because of the structure of the symmetry group, no other operators can transform under $(-1)^F\circ (-\mathds{1}_{G_0})$ or $(-1)^F$ respectively.  
 This example of symmetry fractionalization has a simple interpretation in terms of UV physics. If we think of the line operators as being the world-line of some heavy particle, then the symmetry fractionalization can be thought of as describing a heavy particle that transforms projectively under the 0-form symmetry group.

Let us focus on the case of $Spin_G$-symmetry which is not bosonic. 
 Gluing two copies of $S^2\times D^2$ with the generalized Gluck twist will generate $\ICP^2\#\overline{\ICP}^2$ coupled to a fixed, non-trivial $G_0/\IZ_2$-bundle which has a discrete 2-form flux $w_2(G_0)=w_2(\ICP^2\#\overline{\ICP}^2)$:
 \eq{
 \big\langle D^2\times S^2\big{|}\widehat{\CT}_G\big{|} D^2\times S^2\big\rangle=Z[\ICP^2\#\overline\ICP^2;w_2(G_0)]~.
 }
Now, if the generalized Gluck twist operator acts non-trivially on any line operator $\widehat{L}$ that generates a state in $\CH[S^2\times S^1]$, then $\widehat{L}$ will also necessarily have non-vanishing expectation value on $S^4$, spontaneously breaking the $\IZ_2^{(1)}$ symmetry.  To see this, let us assume there is a state $|\psi\rangle \in \CH\big[S^2\times S^1]$ that transforms non-trivially under $\widehat{\CT}_G$ which is constructed by inserting a charged/fractionalized line operator $\widehat{L}$ in the path integral on $D^3\times S^1:$
\eq{
\widehat{\CT}_G|\psi\rangle=-|\psi\rangle\quad, \quad |\psi\rangle=\widehat{L}\,|D^3\times S^1\rangle~. 
}
We can now compute the inner product 
\eq{
\langle S^2\times D^2|\psi\rangle=\langle S^2\times D^2 |\widehat{L}|D^3\times S^1\rangle=\langle \widehat{L}\rangle_{S^4}=\langle\mathds{1}\rangle_{S^4}=1~,
}
where here we used the fact that $\pi_1(S^4)=0$. This implies that if the generalized Gluck operator $\widehat{\CT}_G$ acts non-trivially on some state on $\CH[S^2\times S^1]$, then the theory will experience symmetry fractionalization with respect to $(-1)^F\circ (-\mathds{1}_{G_0})$.  
In other words, the generalized Gluck twist acts trivially in a theory without $Spin_G$-symmetry fractionalization.

Therefore, in a $4d$ unitary TQFT without $Spin_G$-symmetry fractionalization we have
\eq{
Z_{{\rm Spin}_G}\big[\ICP^2\#\overline\ICP^2;w_2(G_0)\big]=Z_{{\rm Spin}_G}\big[S^2\times S^2\big]\neq 0~,
}
which implies that 
\eq{
Z_{{\rm Spin}_G}\big[\ICP^2;w_2(G_0)\big]\neq 0~.
} 

In the case of bosonic theories, the above results generalize straight-forwardly and the analogous statement is that in a bosonic $4d$ TQFT without $(-1)^F$-symmetry fractionalization
\eq{
Z_{\rm bos}[\ICP^2;0]\neq 0~.
} 

\subsection{Positivity in Higher Dimensions}

Now we are prepared to prove our main result. As in Section \ref{sec:spin}, the partition function of a $d$-dimensional unitary TQFT has a non-trivial partition function on $S^d$: $Z\big[S^d\big]=1$. Since we have already discussed the case for $d=4$, let us now focus on the example of $d=6$. 

Let us now consider the case of $Spin_G$-symmetries. 
Again, by cutting the sphere, along a $S^{3}\times S^{2}$ we can deduce that $Z\big[S^{4}\times S^{2}\big]\neq 0$. We can again cut along $S^2\times S^1\times S^{2}$:
\eq{
S^4\times S^{2}=(\overline{D^3\times S^1\times S^{2}})\#(S^2\times D^2\times S^{2})
}
to see that 
\eq{
Z_{{\rm Spin}_G}\big[S^4\times S^{2}\big]=\big\langle D^3\times S^1\times S^{2}\big{|}S^2\times D^2\times S^{2}\big\rangle\neq 0}
which further implies that $Z_{{\rm Spin}_G}\big[S^2\times S^2\times S^{2}\big]\neq 0$. 

From our discussion of the generalized Gluck twist, we now can see that 
\eq{
Z_{{\rm Spin}_G}\big[(\ICP^2\#\overline{\ICP}^2)\times S^{2};w_2(G_0)\big]=\big\langle D^2\times S^2\times S^{2}\big{|}\widehat{\CT}_G\big{|}D^2\times S^2\times S^{2}\big\rangle
}
where $\widehat{\CT}_G=(-1)^F\circ (-\mathds{1}_{G_0})$. In the case where this acts trivially on the vacuum -- i.e. there is no $Spin_G$-symmetry fractionalization -- we then have the relation:
\eq{
Z_{{\rm Spin}_G}\big[(\ICP^2\#\overline{\ICP}^2)\times S^{2};w_2(G_0)\big]=Z_{{\rm Spin}_G}\big[S^2\times S^2\times S^{2}\big]\neq 0~.
}
We can then use the Proposition \eqref{prop} to deduce
\eq{
Z_{{\rm Spin}_G}\big[(\ICP^2\#\overline{\ICP}^2)\times S^{2};w_2(G_0)\big]&=Z_{{\rm Spin}_G}\big[\ICP^2 \times S^{2};w_2(G_0)\big]\,Z_{{\rm Spin}_G}\big[\overline{\ICP}^2\times S^{2};w_2(G_0)\big]\\
&\neq 0
}
which then implies 
\eq{
Z_{{\rm Spin}_G}\big[\ICP^2\times S^{2};w_2(G_0)\big]\neq 0~.
}
Now, if we further assume that the $G_1$ symmetry is not spontaneously broken, then  we can use the Lemma from Section \ref{sec:lemma} to show that the partition function on $\ICP^2\times S^{2}$ is non-zero when we turn on (additional) background gauge fields $\{A\}$ for $G_1$:
\eq{\label{nonspinnonzero}
Z_{{\rm Spin}_G}\big[\ICP^2\times S^{p+1};w_2(G_0),A\big]\neq0~.
}
The case of bosonic symmetries follows directly from the above result by replacing the generalized Gluck twist with the standard Gluck twist: $\widehat{\CT}_G\mapsto \widehat{\CT}$. As discussed above, this has effect of not turning on the discrete $w_2(G_0)$ flux which leads to the result:
\eq{
Z_{\rm bos}\big[\ICP^2\times S^{2};A\big]\neq 0~.
}

\subsection{Anomaly Enforced Gaplessness for $Spin_G$ Symmetries in General Dimensions}
Now we can analyze a $d$-dimensional QFT that has an anomaly described by the $(d+1)$-dimensional cohomology class $\omega_{d+1}(A)$ where $d=4,6$. Here we use $A$ to denote the background fields as well as dependence on the Stiefel-Whitney classes of $\ICP^2\times S^{p}$ (where $p=d-4$) and denote the corresponding $Spin_G$-structure  by $w_2(G_0)$. Again, the phase that the partition function picks up under a constant background gauge transformation $g\in G$ is identical to the evaluation of $\omega_{d+1}(A)$ on the product manifold $N_{d+1}^{(g)}=(\ICP^2\times S^{p})\rtimes_g S^1$ where we twist the $Spin_G$ bundle along the $S^1$ direction by $g$:
\eq{
Z_{QFT}\big[\ICP^2\times S^{p};A^{(g)}\big]=
{\rm exp}\left\{i \oint_{N_{d+1}^{(g)}}\omega_{d+1}(A)\right\}\times Z_{QFT}\big[\ICP^2\times S^{p};A\big]~. 
}
Again, since $A^{(g)}=A$, we can see that either 
\eq{
{\rm exp}\left\{i \oint_{N_{d+1}^{(g)}}\omega_{d+1}(A)\right\}=1\quad {\rm or}\quad Z_{QFT}\big[\ICP^2\times S^{p};A\big]=0~.
}
If this anomaly is to be matched by a symmetry preserving, unitary TQFT without $Spin_G$ symmetry fractionalization, then \eqref{nonspinnonzero} implies that 
\eq{
{\rm exp}\left\{i \oint_{N_{d+1}^{(g)}}\omega_{d+1}(A)\right\}=1~.
}
This means that if 
\eq{
{\rm exp}\left\{i \oint_{N_{d+1}^{(g)}}\omega_{d+1}(A)\right\}\neq 1
}
then $Z_{QFT}\big[\ICP^2\times S^{p};A\big]=0$ which cannot be matched by a unitary, symmetry preserving TQFT unless it spontaneously breaks $G$ or exhibits $Spin_G$ symmetry fractionalization. This implies that the theory cannot flow in the IR to a symmetry preserving gapped phase (i.e. symmetry preserving unitary TQFT). 
This proves our main result.

 \section{Examples in $4d$}
 \label{sec:example4d}
 
We will now describe some examples of anomalies in $4d$ theories that obstruct symmetry preserving gapped phases. Here we will only present examples for which the constraint arises from the theory on a non-spin manifold. Two such anomalies are given by: 
\eq{ 
\CA_1=\pi i \int x_1\cup y_2\cup w_2(TM)\quad, \quad \CA_2=\pi i \int w_2(TM)\cup z_3~.
}
We will refer to these anomalies as ``Type 1'' and ``Type 2'' $\IZ_2$-anomalies according to their subscript. 

\subsection{Type 1 Anomaly}

Here we will study the Type 1 $\IZ_2$-anomaly. This anomaly describes a mixed anomaly between a 0-form $\IZ_2^{(0)}$-symmetry with background gauge field $x_1$, 1-form $\IZ_2^{(1)}$-symmetry with background gauge field $y_2$, and gravity. 

 This anomaly can be activated on the $5d$ mapping class torus  $N_5=\ICP^2\rtimes_{-\mathds{1}}S^1$ where we activate $y_2,w_2(TM)$ as the generator of $H^2(\ICP^2;\IZ_2)$ and twist the 0-form symmetry by $-\mathds{1}$ along the $S^1$: i.e. pick $x_1$ so that $H^1(S^1;\IZ_2)=\IZ_2[x_1]$. With this choice of background gauge fields we can evaluate
\eq{
{\rm exp}\left\{\pi i \int_{N_5}x_1\cup y_2\cup w_2(TM)\right\}={\rm exp}\left\{\pi i \int_{\ICP^2}w_2(TM)\cup w_2(TM)\right\}=e^{\pi i}=-1~. 
}

This anomaly can be matched by a $4d$ $\IZ_2$-gauge theory with action 
\eq{
S=\pi i \int \big( b_2\cup \beta(a_1)+b_2\cup w_2(TM)+x_1\cup y_2\cup a_1\big)~,
}
where $a_1,b_2$ are dynamical $\IZ_2$-valued 1- and 2-form fields. 
However, this TQFT spontaneously breaks $(-1)^F$ symmetry. This can be seen explicitly by considering the effect of the middle term 
\eq{
S=...+\pi i \int b_2\cup w_2(TM)~.
}
This theory has a single non-trivial line operator and a single non-trivial surface operator. The above term has the effect of making the line operator a fermion which means that $(-1)^F$ has fractionalized.\footnote{Note that another way this can be matched by a $\IZ_2$-gauge theory in the IR is by 
\eq{
S^\prime=\pi i \int \big(b_2\cup \beta(a_1)+b_2\cup y_2+x_1\cup w_2(TM)\cup a_1\big)~,
}
where again $a_1,b_2$ are dynamical $\IZ_2$-gauge fields. However, now the middle term implies that the $a_1$ Wilson line operator is now charged under the 1-form $\IZ_2^{(1)}$ symmetry which spontaneously breaks the symmetry. }

These anomalies arise naturally in the setting of $4d$ QCD-like theories with a 1-form center symmetry and $Spin_G$ structure. In this setting, $U(1)$ 0-form symmetries which act on fermions as a ``flavor symmetry'' are generically broken to a discrete subgroup $G^{(0)}$ due to ABJ anomalies.\footnote{Here we would like to point out that in the more general setting, these symmetries can sometimes be expanded to non-invertible symmetries as discussed in \cite{Cordova:2022ieu,Choi:2022jqy,Choi:2022rfe,Choi:2022fgx}.} When there is an additional 1-form center symmetry $G^{(1)}$, the ABJ anomaly often leads to a mixed anomaly between the remaining 0-form symmetry and the 1-form center symmetry:
\eq{
\CA\sim 2\pi i \frac{p}{q}\int x_1\cup \CP(y_2)
}
where here $x_1$ is the $G^{(0)}$ background gauge field, $\CP$ is the Pontryagin square and $y_2$ is the $G^{(1)}$ background gauge field. Here the fraction $\frac{p}{q}$ depends on the global form of the gauge group and the representations of the fermions. 

When $\frac{p}{q}=\half~{\rm mod}_\IZ$, this anomaly reduces to a $\IZ_2$ anomaly that can only be activated on a non-spin manifold. This is because of the series of mod$_2$ relations 
\eq{
\CP(y_2)=y_2\cup y_2~{\rm mod}_2=y_2\cup w_2(TM)~{\rm mod}_2
}where we used the Wu formula, 
so that 
\eq{
\CA=\pi i \int x_1\cup y_2\cup w_2(TM)
} 
which is of Type 1.

\subsection{Type 2 Anomaly}

Now we will study the Type 2 $\IZ_2$-anomaly. This anomaly can be thought of as a mixed anomaly between gravity and a 1-form $\IZ_2^{(1)}$ global symmetry where $z_3=\beta(y_2)$ is the Bockstein of the $\IZ_2^{(1)}$ background gauge field. 

This anomaly is also activated on the $5d$ mapping class torus $N_5=\ICP^2\rtimes_{-\mathds{1}} S^1$ where we twist the $Spin_G$ bundle by $-\mathds{1}_G$ so that $z_3$ decomposes as 
\eq{
z_3=y_2\cup w_1\quad, \quad y_2\in H^2(\ICP^2;\IZ_2)\quad, \quad w_1\in H^1(S^1;\IZ_2)~. 
}
See \cite{Wang:2014lca,BI} for an explicit construction of mapping class tori that activate such an anomaly. 

Again, this anomaly can be matched in the IR via a $\IZ_2$-gauge theory with actoin 
\eq{
S=\pi i \int\left(b_2\cup \beta(a_1)+b_2\cup w_2(TM)+z_3\cup a_1\right)~.
}
However, here we again see that the TQFT exhibits $(-1)^F$ symmetry fractionalization. 

These types of anomalies also appear in many interesting examples. One of the simplest is $4d$ $SU(2)$ gauge theory with a Weyl fermion in the 4-dimensional representation which has an anomaly where $z_3=w_3(TM)$ \cite{Wang:2018qoy}. This theory admits a symmetry preserving deformation that triggers an RG flow to ``all fermion electrodynamics'' which is a manifestly gapless theory that matches the anomaly. 

Historically, the first instance of this anomaly was in $4d$ $SU(2)$ $\CN=2$ supersymmetric Yang-Mills theory where $z_3=\beta(y_2)$ where $y_2$ is the background gauge field for the $\IZ_2^{(1)}$ center symmetry  \cite{Donaldson:1983wm,Donaldson:1990kn,Witten:1988ze,Witten:1994cg,Moore:1997pc}. As shown by Seiberg-Witten \cite{Seiberg:1994rs,Seiberg:1994aj}, this theory flows to a gapless  supersymmetric $U(1)$ gauge theory. This also has a SUSY breaking, yet symmetry preserving deformation that flows to $4d$ $SU(2)$ non-supersymmetric Yang-Mills theory with two adjoint fermions which is also believed to flow to a gapless $\ICP^1$ non-linear sigma model \cite{Shifman:2013yca,Cordova:2018acb}. Note that in these examples, the theory has a $Spin_{SU(2)}$ structure and $G_1=\IZ_2^{(1)}$.  

Theories with Type 2 anomalies studied recently by \cite{Brennan:2022tyl,BI} in which the authors have constructed robust methods using RG flows to abelian phases and index theorems to probe for these anomalies. With the results presented here, we can further see that any theory which has such an anomaly must either flow to a spontaneously broken gapped phase, symmetry fractionalized gapped phase, or to a gapless theory in the deep IR.

\section*{Acknowledgements}
The author would like to thank   Sungwoo Hong,  Ken Intriligator,  Po-Shen Hsin, Zhengdi Sun, Kantaro Ohmori, and Clay C\'ordova for helpful discussions and related collaborations. 
TDB is supported by Simons Foundation award 568420 (Simons
Investigator) and award 888994 (The Simons Collaboration on Global Categorical Symmetries).

\bibliographystyle{utphys}
\bibliography{SymmetryEnforcedGaplessnessBib}

\end{document}